\documentclass[sigconf,anonymous=False,review=False,natbib=True]{acmart}
\usepackage{tabularx, booktabs}
\usepackage{graphicx}
\usepackage{multirow}
\usepackage{wasysym}
\usepackage{setspace}
\usepackage{xcolor}
\usepackage{array}
\usepackage{balance}
\usepackage{enumitem}
\usepackage{booktabs}
\usepackage{makecell}
\usepackage{geometry}
\usepackage{subcaption}
\usepackage[ruled]{algorithm2e}
\SetKw{kwInit}{Init}
\usepackage{acronym}

\acrodef{IR}{Information Retrieval}
\acrodef{MLP}{multilayer perceptron}
\acrodef{SERP}{Search Engine Result Page}
\acrodef{ERP}{event related potential}
\acrodef{EEG}{electroencephalogram}
\acrodef{DT}{Gradient Boosting Decision Tree}
\acrodef{SST}{SST-EmotionNet}
\acrodef{AUC}{Area Under Curve}
\acrodef{MAE}{Mean Absolute Error}
\acrodef{IN}{Information Need}
\acrodef{fMRI}{functional magnetic resonance imaging}
\acrodef{BCI}{brain–computer interface}
\acrodef{BP}{band power}
\acrodef{DE}{differential entropy}
\acrodef{RASM}{rational asymmetry}
\acrodef{DASM}{differential asymmetry}
\acrodef{SVM}{support vector machines}
\acrodef{DBN}{deep belief networks}
\acrodef{KNN}{k-Nearest Neighbors}
\acrodef{HBG}{Height-Biased Gain}
\acrodef{BTA}{Brain Topography Adaptive Network}
\acrodef{CNN}{convolutional neural network}
\acrodef{GCN}{graph convolutional network}
\acrodef{NDCG}{Normalized Discounted Cumulative Gain}
\acrodef{MAP}{Mean Average Precision}
\acrodef{NLP}{Natural Language Processing}
\acrodef{MI}{Motion Imaginary}
\acrodef{EC}{Emotion Recognition}
\acrodef{RNN}{Recurrent Neural Network}
\acrodef{VR}{Virtual Reality}
\AtBeginDocument{%
  \providecommand\BibTeX{{%
    \normalfont B\kern-0.5em{\scshape i\kern-0.25em b}\kern-0.8em\TeX}}}

\makeatletter
\def\hlinew#1{%
  \noalign{\ifnum0=`}\fi\hrule \@height #1 \futurelet
   \reserved@a\@xhline}
\makeatother

\copyrightyear{2022}
\acmYear{2022}
\setcopyright{acmcopyright}
\acmConference[MM '22] {Proceedings of the 30th ACM International Conference on Multimedia }{October 10--14, 2022}{Lisboa, Portugal.}
\acmBooktitle{Proceedings of the 30th ACM International Conference on Multimedia (MM '22), October 10--14, 2022, Lisboa, Portugal}
\acmPrice{15.00}
\acmISBN{978-1-4503-9203-7/22/10}
\acmDOI{10.1145/3503161.3548258}
\begin{document}

\title{Brain Topography Adaptive Network for Satisfaction Modeling in Interactive Information Access System}

\author{Ziyi Ye}
\email{yeziyi1998@gmail.com}
\affiliation{%
  \institution{BNRist,DCST,Tsinghua University}
  \city{Beijing}
  \country{China}}
 
\author{Xiaohui Xie}
\email{xiexh_thu@163.com}
\affiliation{%
  \institution{BNRist,DCST,Tsinghua University}
  \city{Beijing}
  \country{China}}

\author{Yiqun Liu$*$}
\email{yiqunliu@tsinghua.edu.cn}
\affiliation{%
  \institution{BNRist,DCST,Tsinghua University}
  \city{Beijing}
  \country{China}}
 
\author{Zhihong Wang}
\email{wangzhh629@mail.tsinghua.edu.cn}
\affiliation{%
  \institution{BNRist,DCST,Tsinghua University}
  \city{Beijing}
  \country{China}}

\author{Xuesong Chen}
\email{chenxuesong1128@163.com}
\affiliation{%
  \institution{BNRist,DCST,Tsinghua University}
  \city{Beijing}
  \country{China}}

\author{Min Zhang}
\email{z-m@tsinghua.edu.cn}
\affiliation{%
  \institution{BNRist,DCST,Tsinghua University}
  \city{Beijing}
  \country{China}}

\author{Shaoping Ma}
\email{msp@tsinghua.edu.cn}
\affiliation{%
  \institution{BNRist,DCST,Tsinghua University}
  \city{Beijing}
  \country{China}}

\thanks{$*$Yiqun Liu is the corresponding author.}

\renewcommand{\shortauthors}{Ziyi Ye et al.}

\begin{abstract}
With the growth of information on the Web, most users heavily rely on information access systems~(e.g., search engines, recommender systems, etc.) in their daily lives.  
During this procedure, modeling users' satisfaction status plays an essential part in improving their experiences with the systems.
In this paper, we aim to explore the benefits of using Electroencephalography~(EEG) signals for satisfaction modeling in interactive information access system design.
Different from existing EEG classification tasks, the arisen of satisfaction involves multiple brain functions, such as arousal, prototypicality, and appraisals, which are related to different brain topographical areas.
Thus modeling user satisfaction raises great challenges to existing solutions. 
To address this challenge, we propose BTA, a \textbf{B}rain \textbf{T}opography \textbf{A}daptive network with a multi-centrality encoding module and a spatial attention mechanism module to capture cognitive connectives in different spatial distances.
We explore the effectiveness of BTA for satisfaction modeling in two popular information access scenarios, i.e., search and recommendation.
Extensive experiments on two real-world datasets verify the effectiveness of introducing brain topography adaptive strategy in satisfaction modeling.
Furthermore, we also conduct search result re-ranking task and video rating prediction task based on the satisfaction inferred from brain signals on search and recommendation scenarios, respectively.
Experimental results show that brain signals extracted with BTA help improve the performance of interactive information access systems significantly.

\end{abstract}



\maketitle

\section{Introduction}

Web-based information access systems~(e.g., search engines and recommender systems) heavily rely on implicit feedback~(e.g., clicks, dwell time, and eye-tracking) to improve user experience~\cite{fox2005evaluating, wu2019investigating}.
However, implicit feedback acts as an indirect probe of user feelings and thus is sometimes biased and misleading~\cite{amatriain2009like, wang2021clicks, liu2014skimming}.
Therefore, advancement is still required with novel user signals to model the information access process.
In recent years, the developments of \ac{EEG} devices make it feasible to collect brain signals in almost real-time.
\ac{EEG} directly captures brain activities, which can potentially reveal the true underlying user satisfaction while users are accessing information on the Web.
Based on the estimated user satisfaction, the user's intent or preference can be better understood and successfully used to improve user experience.

\begin{figure}[h]
\label{fig:search}
  \centering
  \includegraphics[width=1\linewidth]{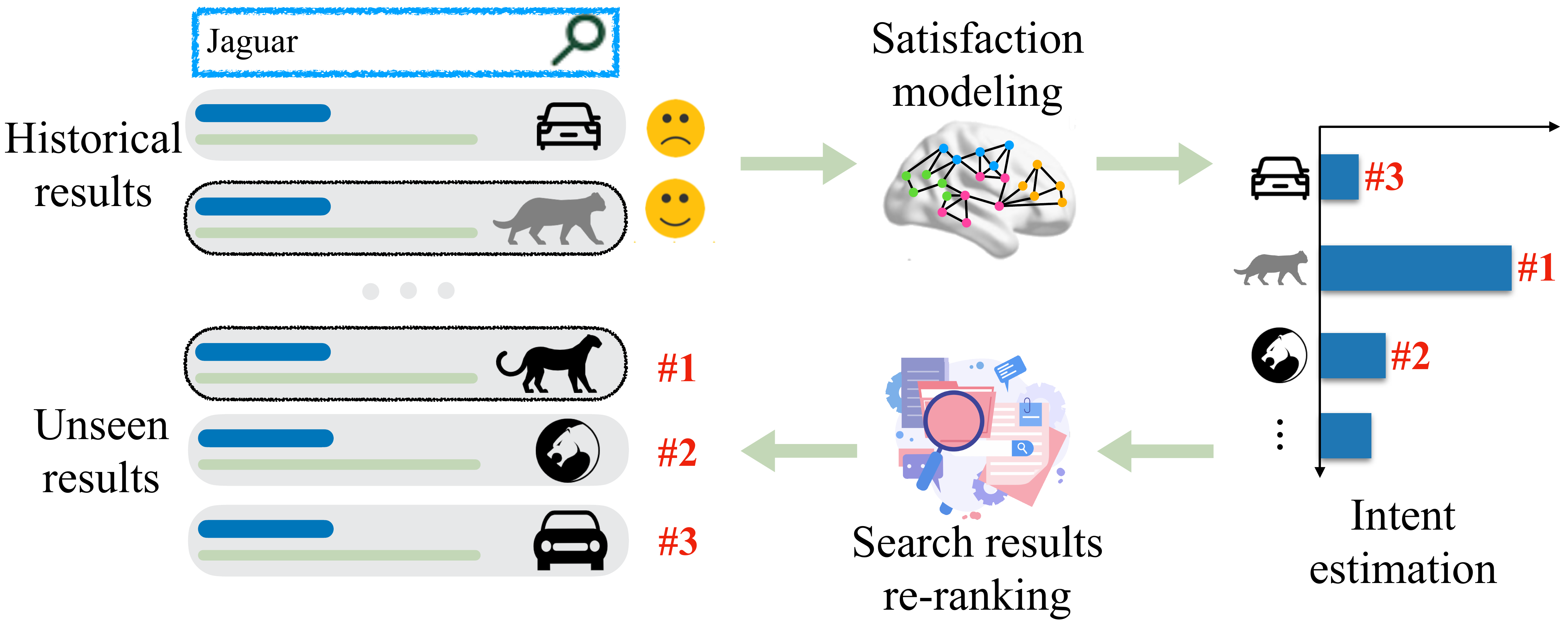}
  \caption{An example of EEG-enhanced search engine. By modeling user satisfaction, we interactively predict user intent and provide potentially more satisfactory results.} 
  \label{fig:search_introduction}
  \Description[]{}
\end{figure}

However, to the best of our knowledge, few studies have thoroughly investigated the methods of utilizing brain signals for satisfaction modeling in the information access procedure. 
Little is known about to what extent search and recommendation system can be benefited by brain signals. 
In Figure~\ref{fig:search_introduction}, we provide a possible interactive system that re-ranks the search results upon the real-time satisfaction modeling.
A user examines two search results corresponding to the query ``Jaguar'', and our system estimates the user satisfaction with brain signals.
Based on the satisfaction modeling, we reestimate the user intent and find that the user intent is related to the ``Jaguar animal''.
Then unseen search results~(i.e., search results on the next screen/page) are re-ranked, and the ones with the subtopic of ``Jaguar animal'' are placed higher than those of ``Jaguar browser'' and ``Jaguar car''.

Different from existing \ac{EEG} classification tasks, modeling user satisfaction is challenging since satisfaction involve both cognitive processes~(e.g., prototypicality~\cite{martindale1988priming} and appraisals~\cite{silvia2005cognitive}) and affective states~(e.g., arousal~\cite{berlyne1970novelty}).
This combined process starts at the occipital region and engages frontal-parietal attentional circuits, which involve brain functions in dorsolateral frontal region and medial temporal region~\cite{chatterjee2003prospects,nadal2008towards}. 
To connect and aggregate the signals in various brain regions, we believe capturing the topographical relations of brain signals is necessary in satisfaction modeling.   


Existing studies have achieved many compelling outcomes by introducing the topographical information into \ac{EEG} classification models.
Among these models, \ac{CNN}~\cite{lawhern2018eegnet,kostas2021bendr} or \ac{GCN}~\cite{song2018eeg,zhong2020eeg,jia2021hetemotionnet,li2021multi} are most widely used.
They have achieved high classification performance in motion imaginary~\cite{lawhern2018eegnet} and emotion recognition tasks~\cite{li2021multi}.
However, most of them fail to capture the adaptive topography connections.
For example, EEGNet~\cite{lawhern2018eegnet} utilizes convolutions in local brain regions yet ignores the ulterior topographical relations.
DGCNN~\cite{song2018eeg} and RGNN~\cite{zhong2020eeg} apply a public adjacency matrix for graph convolutions, and thus the topography connections are fixed among different data samples.
Since satisfaction is associated with several cognitive processes and affective states, the topographical relations of brain signals is changing and sophisticated. 
Thus it is challenging yet valuable to capture the topographical relations in a data-dependent way. 

To tackle above challenges, we design a novel architecture for \ac{EEG}-based satisfaction modeling named Brain Topography Adaptive network~(BTA).
The network applies a multi-centrality encoding module to generate encodings with 3D topographical information.
Then, motivated by the success of the attention mechanism~\cite{vaswani2017attention}, we adopt a spatial attention module to capture the brain's cognitive connectivity in a data-dependent manner.
The effectiveness of BTA is verified in two typical interactive information access scenarios,  i.e., search and recommendation.
Experimental results in two public datasets~(i.e., one for search and one for recommendation) demonstrate that BTA outperforms various baselines in terms of satisfaction estimation.
Furthermore, we explore to what extent we can utilize the satisfaction estimated with brain signals to improve search result re-ranking and video rating prediction performance for search and recommendation scenarios, respectively.
We demonstrate that the performance can be significantly improved with the satisfaction inferred from brain signals.

In summary, our contributions are three-fold. 
1)~We identify and tackle a novel problem of satisfaction modeling for interactive information access systems using brain signals.
2)~We propose the Brain Topography Adaptive Network~(BTA), which adaptively captures the topographical information to solve the \ac{EEG}-based satisfaction prediction problem. 
Empirical experiments show that our method outperforms state-of-the-art \ac{EEG} classification models.
3)~We explore the possible advancements in interactive information systems brought by the estimated user satisfaction inferred from brain signals.
As far as we know, this is the first time \ac{EEG}-based satisfaction modeling is successfully applied to interactive search~(i.e., search result re-ranking) and recommendation~(i.e., video rating prediction) tasks.
 



\section{related work}
\subsection{User Satisfaction in Information Access}
Satisfaction measures users’ subjective feelings about the system, which can be treated as the fulfillment of their information requirement~\cite{kelly2009methods}.
It has been noticed that modeling user satisfaction is valuable for performance improvement and evaluation in information systems~\cite{ali2011overview,liu2018satisfaction}.

Researchers have focused on modeling users’ satisfaction with users’ implicit feedback signals~(click, dwell time, scroll, etc.)~\cite{fox2005evaluating, hassan2014struggling}.
Recent years have witnessed much research introducing novel additional user signals and corresponding strategies to estimate user satisfaction, e.g., mouse movements~\cite{chen2017user} and eye-tracking~\cite{wu2019investigating}. 
However, implicit feedback signals are just indirect probes of real user satisfaction and thus are often incorrect~\cite{amatriain2009like}.
In search scenarios, \citet{liu2014skimming} reveal that a large proportion~(45.8\%) of eye fixations are irrelevant to relevance estimation.
In news recommendation, \citet{wang2021clicks} find that users' click behaviors can not be simply treated as positive signals due to the ``clickbait'' effect.

In this study, we explore satisfaction modeling with brain signals with our proposed BTA. 
Besides, we conduct search and recommendation tasks with the satisfaction inferred from brain signals.
We aim to demonstrate the benefits of brain signals as effective user feedback for designing an interactive information access system.

\subsection{Neuroscience \& Information Access System}
Recent research has applied brain imaging tools to study aspects of the information access process from a neuroscience perspective.
For example, \citet{moshfeghi2016understanding,moshfeghi2019towards} conduct a series of studies to unravel the neural basis of information need.
Apart from neurological-based analysis, researchers also utilize brain signals to predict user judgments in the information access process.
For example, \citet{gwizdka2017temporal,kim2019erp} conduct extensive studies to judge text relevance and topical relevance of visual shots using \ac{EEG} signals.
\citet{davis2021collaborative} conduct an interesting study to estimate facial preference with brain signals and then utilizes the estimated preference for collaborative filtering.
However, few studies have thoroughly investigated the satisfaction modeling in information systems and explored to what extent the system can be improved with it. 

What we add on top of prior works is that we design special \ac{EEG} classification models and conduct interactive information access tasks to demonstrate the benefits of introducing brain signals into user satisfaction modeling.

\subsection{EEG-based Classification}
\ac{EEG} has the advantages of high temporal resolution, non-invasiveness, and relatively low financial cost.
It is widely used in researches involving motion imaginary~\cite{lawhern2018eegnet, kostas2021bendr}, emotion recognition~\cite{jia2021hetemotionnet,li2021multi,zhong2020eeg}, and sleep stage scoring~\cite{jia2020graphsleepnet,supratak2020tinysleepnet}.
Recently trends of EEG-based classification have been developed substantially from topology-invariant algorithms~(e.g., \ac{SVM}~\cite{wang2014emotional}, \ac{DT}~\cite{guan2019motor}, and \ac{RNN}~\cite{li2016emotion}) to topology-aware methods such as \ac{CNN} and \ac{GCN}.
The topology-aware models take the topological structure of \ac{EEG} features into account when learning the representations.
For example, \citet{lawhern2018eegnet} and \citet{kostas2021bendr} map the 3D \ac{EEG} topographical information into 2D representations and adopt \ac{CNN}-based architecture to aggregate channel informations.
Additionally, there are some studies~\cite{zhong2020eeg, song2018eeg, zhang2021sparsedgcnn} applying \ac{GCN}-based methods, and a public adjacency matrix is adopted to automatically learn the aggregation weight between \ac{EEG} channels.
And \citet{li2021multi} exploit multi-domain information to build the adjacency matrix for their \ac{GCN}-based model. 

However, as satisfaction is related to various cognitive processes and affective states~\cite{martindale1988priming, silvia2005cognitive, berlyne1970novelty}, the brain connectivities might be different due to different stimulus factors~\cite{ho2015emotion,yin2017functional}.
Few studies adaptively learn the brain connectivities in a completely data-dependent way.
One exception is the HetEmotionNet~\cite{jia2021hetemotionnet}, which utilizes mutual information~\cite{kraskov2004estimating} to construct the adjacency matrix for each data sample and then applies the graph convolution with it.
However, the mutual information indicates the similarity of \ac{EEG} features.
Simply aggregating \ac{EEG} channels with similar features is not always reasonable.
For example, previous studies have suggested that the asymmetry in neuronal activities between the left and right hemispheres is informative~\cite{schmidt2001frontal}.  
Our study differs from previous work in using a completely data-dependent aggregation strategy with the specially designed multi-centrality encoding and spatial attention mechanism.


\section{EEG-based Satisfaction Modeling}
\subsection{Problem statement}
In this paper, we define the temporal features of \ac{EEG} signals as $X^{t}=\{x^t_1, x^t_2, ..., x^t_N\} \in \mathbb{R}^{N \times E}$,  where $N$ is the length of temporal features, $E$ is the number of \ac{EEG} channels.
The spectral features of \ac{EEG} signals are denoted as $X^{s}=\{x^s_1, x^s_2, ..., x^s_B\} \in \mathbb{R}^{B \times E}$, where $B$ is the length of spectral features, $E$ is the number of \ac{EEG} channels.
The spectral features are the \ac{DE}~\cite{duan2013differential} extracted from $B$ frequency bands~(e.g., $\delta, \theta, \alpha, \beta$, and $\gamma$).
Then the \ac{EEG}-based satisfaction modeling problem is to learn a mapping function $F_s$ based on the proposed model, which can be formulated as:
$$
\hat{y}=F_s(X^{t}, X^{s})
$$
where $\hat{y}$ denotes the estimated satisfaction score.
With the predicted satisfaction score on historical items, information system can interactively provide more satisfying unseen items.
In this paper, we explore two common information access scenarios, i.e., search and recommendation, which are detailed in Section~\ref{Experimental setup}.

\subsection{Brain Topography Adaptive Network}
\label{Methdology}
\subsubsection{Model overview}
\begin{figure}[t]
  \centering
  \includegraphics[width=1\linewidth]{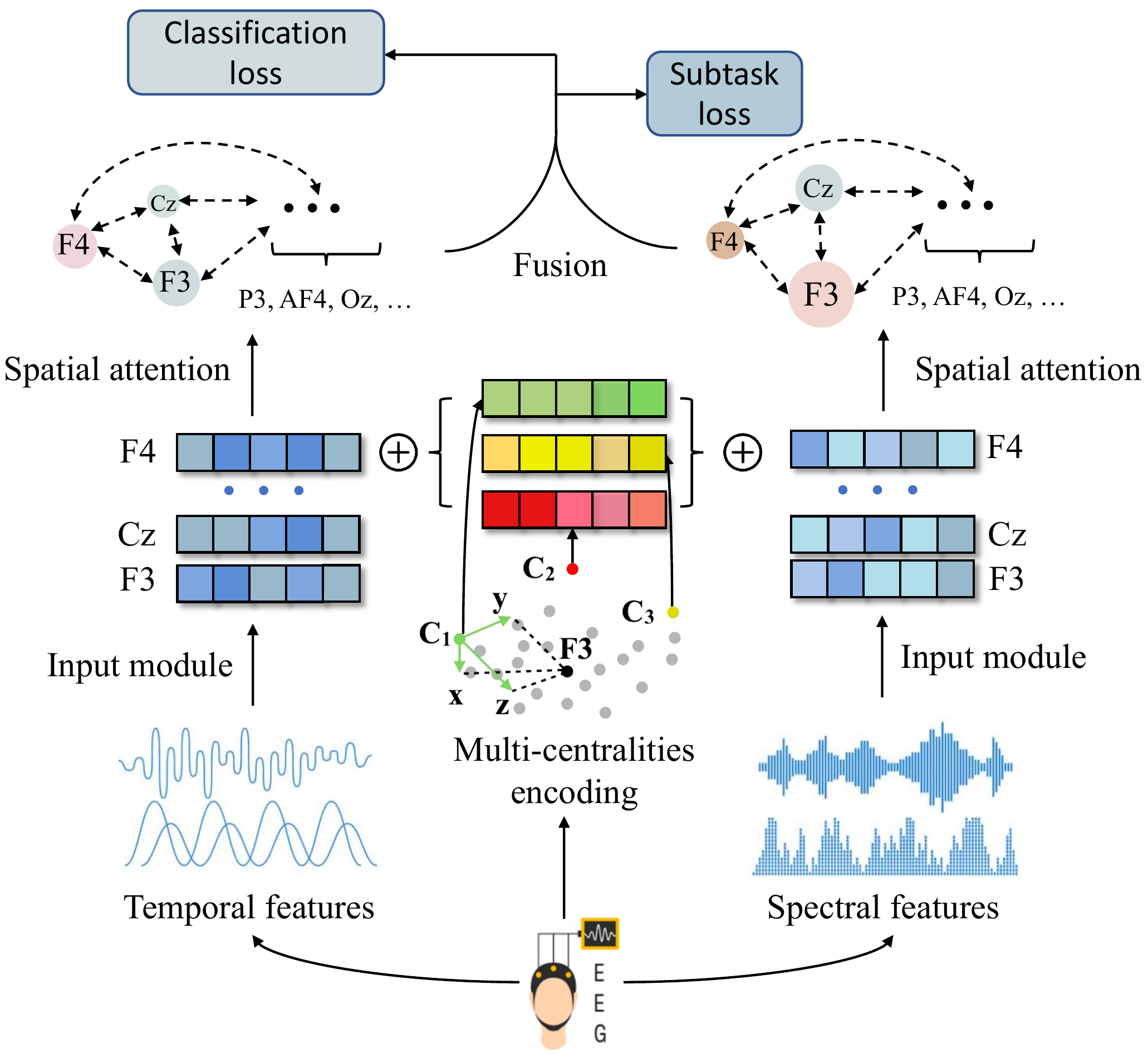}
  \caption{The architecture of proposed BTA. 
  } 
  \label{fig:model}
  \Description[]{}
\end{figure}

An overview of the proposed satisfaction modeling method is illustrated in Figure~\ref{fig:model}. 
We propose a Brain Topography Adaptive Network~(BTA), which consists of a temporal data stream and a spectral data stream, and these two data streams have similar structures.
The model is composed of five components: (1)~\textit{Input encoding module}. (2)~\textit{Multi-centrality encoding module}. (3)~\textit{Spatial attention module}. (4)~\textit{Fusion and classification module}. (5)~\textit{Subtask module}.
Firstly, the spectral features or temporal features are linearly projected into a latent space with the \textit{input module}.
Then, the \textit{multi-centrality encoding module} generates a spatial encoding for each channel according to its topographical relation to various spatial centralities.
After that, the \textit{spatial attention module} aggregates the channel information with attention mechanisms to capture the brain's cognitive connectivity.
Finally, the \textit{fusion and classification module} fuses the temporal data stream and spectral data stream to classify user satisfaction in information access tasks.
Additionally, to initialize the centrality embedding vectors, we replace the \textit{fusion and classification module} with a \textit{subtask module} before the model training.

\subsubsection{Input encoding module}
The spectral features or temporal features are linearly projected into a latent space, which can be formulated as:
$$
H^t = W^t * X^t + B^t, H^s = W^s * X^s + B^s
$$
where $X^t\in \mathbb{R}^{E\times N}, X^s \in \mathbb{R}^{B\times N}$ are temporal and spectral EEG features, respectively, $W^t \in \mathbb{R}^{H\times N}$, $W^s \in \mathbb{R}^{H\times B}$, $B^t \in \mathbb{R}^{H\times E}, B^s \in \mathbb{R}^{H\times E}$ are 
learnable parameters, $H^t \in \mathbb{R}^{H\times E}, H^s \in \mathbb{R}^{H\times E}$ are the input vectors.
We transform $X^t$ and $X^s$ into latent spaces with the same dimension since the multi-centrality encoding module, which adds topographical information to the input vectors~(see in Section~\ref{Multi-centrality encoding module}), is shared between them.
 
 For simplicity, in Section~\ref{Multi-centrality encoding module} and Section~\ref{Spatial attention}, we omit the subscripts $t$ and $s$~(e.g., $H \in \{H^t, H^s\}$) since the temporal and spectral data stream share the same architectures.

\subsubsection{Multi-centrality encoding module}
\label{Multi-centrality encoding module}
To map each channel with a meaningful spatial encoding, we adopt multi-centrality encoding as an additional signal to the neural network.
Specifically, we select $M$ spatial points $C_1,C_2,...,C_M$ in the brain topography space as the centralities. 
For each centrality $C_j$, we build a spherical coordinate system $\mathscr{F}_{j}$ with $C_j$ being regraded as the origin.
And the zenith direction and the azimuth direction of $\mathscr{F}_{j}$ are defined as straight above and ahead of the human brain, respectively.
Then, we develop a centrality encoding $p_{i,j}$ which assigns each \ac{EEG} channel $E_i$ three learnable embedding vectors according to its spherical coordinate in $\mathscr{F}_{j}$.
The centrality encoding $p_{i,j} \in \mathbb{R}^{H}$ of a \ac{EEG} channel $E_i$ according to a centrality $C_j$ is formulated as:
$$
	p_{i,j} = \rho_{i,j} * c_{j, \rho} + \theta_{i,j} * c_{j, \theta} + \varphi_{i,j} * c_{j, \varphi} 
$$
where $(\rho_{i,j}, \theta_{i,j}, \varphi_{i,j})\in \mathbb{R}^{3}$ is the spherical coordinate of $E_i$ in $\mathscr{F}_{j}$, $( c_{j, \rho}, c_{j, \theta}$, $c_{j, \varphi}) \in \mathbb{R}^{3\times H}$ are the centrality embedding vectors related to $C_{j}$. 
To enrich the spatial representation of \ac{EEG} channels and increase their distinction, we obtain $M$ centrality encodings for an \ac{EEG} channel $E_i$, i.e., $p_{i,j}, j \in\{1,...,M\}$.
The centrality encodings are then combined with the input vector by:
$$
z_{i} = h_{i} \oplus \sum_{1 \leq j \leq M} p_{i,j}
$$
where $h_i \in \mathbb{R}^{H}$ is the subvector of $H^\intercal=\{h_1,...,h_e\}$, $H^\intercal$ is the  transposed matrix of $H\in \{H^t, H^s\}$, $\oplus$ is an element-wise operator.
Then the input vector after adding the multi-centrality encoding can be expressed as $Z=\{z_1,...,z_e\}^\intercal \in \mathbb{R}^{E\times H}$.
We adopt a direct addition as $\oplus$ since it is simple to implement and adds minimal overhead to training time and model size. 
We also tried more complex interaction operators, but the results are similar, so we omit them.
By using the multi-centrality encoding in the input, the input encoding vectors obtain additional topographical information. 
Therefore, the model can capture both the topographical correlation and the spectral or temporal information.

\subsubsection{Spatial attention module}
\label{Spatial attention}

The brain's cognitive connectivity between \ac{EEG} channels should not be neglected in \ac{EEG}-based classification tasks.
Therefore, we adopt attention mechanisms to capture the adaptive channel correlations for each data sample.
We first apply a multihead attention layer to calculate the interacted sequence:
\begin{equation}\nonumber
\begin{aligned}
Z_1&=MultiHead({Z}^\intercal, {Z}^\intercal, {Z}^\intercal)\\
&=Concat(head_1 ,head_2 , ...,head_D )W^O
\end{aligned}
\end{equation}
where $head_i = Attention({Z}^{\intercal}W^Q , {Z}^{\intercal}W^K , {Z}^{\intercal}W^V )$, $W^Q, W^K, W^V \in \mathbb{R}^{H\times H/D}$ are trainable parameters, $D$ is the number of heads, and $Attention(Q, K, V)$ is the scaled dot-product attention mechanism~\cite{vaswani2017attention}.
$Concat(\cdot)$ denotes the concatenation operation in the second dimension and the concatenated vector is fed into a linear matrix $W^O \in \mathbb{R}^{H\times H}$ to obtain the spatial interacted vectors $Z_1 \in \mathbb{R}^{E\times H}$.

Next, we apply a batch normalization layer $BN$ to accelerate the training procedure and obtain the output vector $Z_2=BN(Z_1) \in \mathbb{R}^{E\times H}$.
In spite of using the layer normalization as Transformer~\cite{vaswani2017attention} does, we adopt batch normalization since it performs better than layer normalization.
We suggest that using batch normalization can mitigate the effect of instability and outline values related to \ac{EEG} signals~\cite{zerveas2021transformer}.

\subsubsection{Fusion and classification module}
With the above modules, the input temporal features $X_t$ and spectral features $X_s$ are transformed into output vectors $Z^t_{2}$ and $Z^s_{2}$, respectively.
We concatenate the output vectors to obtain $Z_3 = Concat(z_{2,1},...,z_{2,e}) \in \mathbb{R}^{EH}$, where $Z_2 = \{z_{2,1},...,z_{2,e}\} \in \mathbb{R}^{E\times H}$, $Z_2 \in \{Z^t_2, Z^s_2\}$, and $Z_3 \in \{Z^t_3, Z^s_3\}$.
Then to fuse $Z^t_{3}$ and $Z^s_{3}$, we adopt a fully connected layer, an activation function, and a softmax function, which can be formulated as:
$$
\hat{y} = softmax(W\sigma(Concat(Z^t_{3}, Z^s_{3}))+B)
$$ 
where $\sigma$ denotes as activation function~($Gelu$ in our experiments), $W\in  \mathbb{R}^{EH\times 1}$ and $B\in \mathbb{R}^{1}$ are trainable parameters, $\hat{y}$ is the estimated satisfaction score.
Finally, the classification cross entropy is used as the loss function, which is defined as follows:
$$
L=-y\lg(\hat{y})-(1-y)\lg(1-\hat{y})
$$
where $y$ is the true label.

\subsubsection{Subtask module}
Besides, to initialize the centrality embedding vectors $c_{j, \rho}$, $c_{j, \theta}$, and $c_{j, \varphi}$ for each $j$ in $\{1,2,...M\}$, we adopt an unsupervised subtask prior to the satisfaction modeling task.
The unsupervised subtask is a reconstruction task to predict some mask \ac{EEG} features, which share the same architecture except for the classification layer.  
Specially, randomized binary noise masks $W_{t,mask} \in \mathbb{R}^{H\times N}$ and $W_{s,mask} \in \mathbb{R}^{H\times N}$ are generated for each data sample, and the input $X_t$ and $X_s$ are masked by $\widetilde{X} = W_{mask} \odot X, X \in \{X_t, X_s\}$.
Then, we replace $X$ with $\hat{X}$ to generate the output vector $\hat{Z}_2$ with BTA. 
After that, $X'$ is reconstructed with $\hat{Z}_2$ using a linear connection layer: $X'=linear(\hat{Z}_2)$.  
Finally, we utilize the mean squared error loss to present the reconstruction loss, which can be formulated as:
$$
L_{MSE}=\sum_{(i,j)\in Q}{(X'(i,j)-X(i,j))^2}
$$
where we utilize $Q=\{(i,j)| W_{mask}(i,j)=0\}$ to obtain the loss related to the predictions on the masked values.
After the subtask, the centrality vectors of $c_{j, \rho}$, $c_{j, \theta}$, and $c_{j, \varphi}$ for each $j$ in $\{1,2,...M\}$ are set as the initialized parameters for the satisfaction modeling task.
The training procedures of our proposed BTA model are summarized in Algorithm~\ref{algorithm1}.
Note that the unsupervised subtask aims to initialize a better mutli-centrality encoding which reflects the spatial relations of \ac{EEG} channels.
We find that the initialization process is effective when supervised labels are limited while not when using a larger dataset, which is elaborated in Section~\ref{Satisfaction Prediction}.

\section{Experimental setup}
\label{Experimental setup}
This section details the datasets and experimental settings.
The Search-Brainwave~\cite{sigir2022ye} dataset and the AMIGOS~\cite{miranda2018amigos} dataset are applied for the satisfaction modeling experiments~(detailed in~\ref{Dataset}).
Then, we explore the performance of \ac{EEG}-enhanced interactive search and recommendation systems with a search result re-ranking task on the Search-Brainwave dataset and a rating prediction task on the AMIGOS dataset, respectively. 
The implementation code of our experiment is based on PyTorch~\footnote{https://pytorch.org/} and is publicly available in https://github.com/YeZiyi1998/DL4EEG-Classification.

\subsection{Satisfaction prediction}

\subsubsection{Baselines}
\label{baselines}
We exploit three types of \ac{EEG} classification models as baselines, topology-invariant models, \ac{CNN}-based models, and \ac{GCN}-based models.

Topology-invariant models include \ac{SVM}~\cite{suykens1999least}, \ac{DT}~\cite{safavian1991survey}, and \ac{MLP}~\cite{davis2021collaborative}.
We implement these models with the scikit-learn library~\cite{pedregosa2011scikit}.

\ac{CNN}-based models include EEGNet~\cite{lawhern2018eegnet} and the recently proposed BENDR~\cite{kostas2021bendr}.
EEGNet stacks several \ac{CNN} layers to boost its performance.
BENDR applies \ac{CNN} to extract \ac{EEG} features and then uses transformers to capture temporal patterns of \ac{EEG} signals, which achieves state-of-the-art performances in various \ac{BCI} tasks.
It also adopts a cross-dataset pre-training task to initialize the parameters.
We implement EEGNet and BENDR with their open-sourced code and use the public pre-trained weights to initialize the BENDR model.

\ac{GCN}-based models include DGCNN~\cite{song2018eeg}, RGNN~\cite{zhong2020eeg}, and Het-EmotionNet~\cite{jia2021hetemotionnet}.
DGCNN applies a public adjacency matrix to aggregate multichannel EEG information dynamically.
RGNN uses two regularizers to improve the robustness.
Het-EmotionNet uses mutual information to model the topographical information and fuse the temporal and spectral information together.
Here we implement RGNN and Het-EmotionNet with open-sourced codes and DGCNN by ourselves since their code is not available.

\subsubsection{Parameter Setups}
We train BTA with the Adam optimizer~\cite{kingma2014adam}.
The centralities’ number $M$ is selected from $\{1, 3, 5, 7, 9, 14\}$.
Results show that 3 is a proper setting and achieves the best performance.
Besides, we find that selecting centralities in different spatial positions~(e.g., randomly selecting three channels as the centralities) leads to only a marginal difference.
Therefore, we choose these three centralities~(i.e., $C_1$, $C_2$, $C_3$) for simplicity and representative of the spatial distribution of EEG channels. 
Among them, $C_1$ is selected as a center point in the international 10-20 EEG system. $C_2$ and $C_3$ are selected as the left and right mastoid points since they are widely used as the reference channels in existing EEG studies~\cite{yao2019reference}. 
The initialize learning rate, min-batch size, and the hidden dimension $H$~(as denoted in Section~\ref{Methdology}) are tuned and selected from $\{0.01, 0.05\}$, $\{8, 32\}$ and $\{8,16,32\}$, respectively.
The head number in the multihead attention mechanism is set to eight.
Besides, in the unsupervised subtask, we set the random mask ratio as 15\%, the same as prior work in time sequence prediction~\cite{zerveas2021transformer}.

\subsubsection{Protocols}
We apply \ac{AUC} and F1-score as evaluation metrics in our experiments.
For Search-Brainwave dataset, we conduct experiments on each subject and evaluate the models with a \textit{task-independent ten-fold cross validation}:
search tasks are partitioned into ten folds, and we leave each fold for evaluation after training with the remaining folds.
For AMIGOS dataset, following existing studies~\cite{gjoreski2018inter,hu2021scalingnet}, we randomly split the data and apply a \textit{ten-fold cross validation} for each subject. 

\subsection{Downstream task1: Search result re-ranking}

\subsubsection{Task definition}
\label{problem:Search Result Re-ranking}
In the search result re-ranking task, unseen search results are re-ranked by the estimated satisfaction of historical search results. 
We present a search task within a query $Q$ whose search results is a list $D = \langle d^1, ... ,d^N \rangle$. 
We assume that a user historically examines the top several results $D_{hi}=\langle d^1, ... ,d^{N_{hi}} \rangle$, hence the interaction on the $i$-th search result is formulated as $I_i=\{\hat{y_i}, d_i\}$, where $\hat{y_i}$ is the estimated satisfaction score to search result $d_i$.
Then we would like to place the relevant search results in the unseen search results list $D_{un}=\langle d^{N_{hi}+1}, ... ,d^{N} \rangle$ as high as possible.
Therefore, the goal of the search result re-ranking task is:
$$
\max \sum_{Q}{\pi(\hat{D}_{un}, R_{un})}, \hat{D}_{un} = F_{search}(Q, D_{un}, I_1, ..., I_{N_{hi}})
$$
where $\pi$ denotes the evaluation metric, e.g., \ac{NDCG}~\cite{jarvelin2017ir} or \ac{MAP}~\cite{zhu2004recall}, $R_{un}$ is the true relevance label of $D_{un}$, $\hat{D}_{un}$ is the returned re-ranked list, $F_{search}$ denotes the re-ranking strategy that re-ranks the unseen search results according to historical interactions, which is elaborated in Section~\ref{Methods_Task1}.

\subsubsection{Methods}
\label{Methods_Task1}
To generate a better re-ranked list for unseen results $\hat{D}_{un}$, we build a language model~(LM)~\cite{lavrenko2017relevance} and rewrite the query for a better intent estimation.
The language model estimates the relevance of the words in historical search results $d_i=\{w_{i,1},w_{i,2},...\} \in D_{hi}$ to find the most relevant words related to the query $Q=\{q_1,...,q_k\}$. 
Then, the query is rewritten with $l$ most relevant words as $Q^{'}=\{q_1,...,q_k,w_1,...,w_l\}$.
In the language model, the word relevance $R(w)$ can be formulated as:
$$
R(w) = LM(D_{hi}, Q, F_{D_{hi}})
$$ 
where $F_{D_{hi}}$ is a probability distribution of $D_{hi}$.
Generally, a word will have higher relevance if it appears in search results with higher distribution probabilities.
Convention language model sets $F_{D_{hi}}$ as a uniform distribution since we have no prior knowledge, which can be denoted as \textit{uniformed language model}~(ULM).
With the estimated satisfaction inferred from brain signals, we adjust $F_{D_{hi}}$ to assign a higher probability $P(d|D_{hi})$ for satisfying search results:
$$
P(d|D_{hi})=(\lambda+\hat{y}(d))/(\lambda|D_{hi}|+\sum_{d\in D_{hi}}\hat{y}(d))
$$ 
where $\lambda$ is a hyperparameter to smooth the estimated satisfaction score $\hat{y}(d)$.
Here we denote our method as \textit{satisfaction-enhanced language model}~(SLM).
After extending the query $Q$ to $Q^{'}$ with the language model, we adopt ranking algorithms~(e.g., BM25~\cite{robertson2009probabilistic}) with $Q^{'}$ to generate the re-ranked search results list $\hat{D}_{un}$.

In our experiment, we use BM25~\cite{robertson2009probabilistic} as the ranking algorithm and exploit the performance of the original ranking list generated by BM25 and the re-ranked lists with ULM or SLM rewriting the query.
For parameter settings, we empirically set the length of rewritten words $l$ and the hyperparameter $\lambda$ as 5 and 2, respectively.

\subsubsection{Protocols}
For each search task in the Search-Brainwave dataset, if the user examines $cu$ search results, we will re-rank the remanent $N-cu$ search results and evaluate the re-ranked list with the true relevance label of search results.
Note that $cu$ varies with search tasks since the user can stop her search at any time.
Finally, ranking evaluation metrics \ac{NDCG} and \ac{MAP} are adopted to evaluate the averaged performance among all search tasks.

\subsection{Downstream task2: Rating prediction}

\subsubsection{Task definition}
\label{problem:Rating Prediction}
In the rating prediction task, we predict the ratings of unseen user-item pairs based on historical user-item interactions.
Specially, we denote the user and item set as $U$ and $V$, respectively.  
The historical user-item interactions are denoted as $I_{hi}=\{(u,v,\hat{y}_{u,v})|u \in U, v \in V\}$, where $\hat{y}_{u,v}$ is the satisfaction score of item $v$ for user $u$ estimated with brain signals. 
Note that in conventional recommendation tasks, $\hat{y}_{u,v}$ is replaced by explicit ratings or implicit feedback~(e.g., click, liked).
Then the rating prediction task is to better estimate the true label $y_{u,v}$ in the unseen user-item pairs $I_{un}=\{(u,v,y_{u,v})|u \in U, v \in V\}$.
Therefore, the task is formulated as:
$$
\max \pi(I_{un}, F_{rec}(I_{hi}, \{(u,v)|(u,v,y_{u,v}) \in I_{un}\}))
$$
where $\pi$ denotes the evaluation metric, i.e., \ac{AUC} , $F_{rec}$ is the rating prediction strategy elaborated in Section~\ref{Methods_Task2}, and we use the explicit annotation as true label $y_{u,v}$.

\subsubsection{Methods}
\label{Methods_Task2}
The personalized recommendation aims to learn a mapping function $F$ that maps the rating of a given user-item pair with their embeddings: 
$$
y_{(u,v)}=F(e_{u},e_{v})
$$
where $e_{u}, e_{v}$ indicate the user embeddings and item embeddings, respectively, $y_{(u,v)}$ is the rating of the user-item pair $(u,v)$.
For a recommendation system, the true labels $y_{(u,v)}$ are difficult to obtain~\cite{wang2021clicks}. 
Therefore, in the training process, we hypothesize that we can obtain the true labels $y_{(u,v)}$ for a $\alpha$ ratio of user-item pairs.
And we use the estimated satisfaction score $\hat{y}_{(u,v)}$ for other user-item pairs to introduce more training data. 
Finally, the true label $y_{(u,v)}$ is applied to evaluate the model performance. 

We denote function $F$ using $\alpha$ ratio of true labels $T$ as $F^{T(\alpha)}$, and $F$ using $\alpha$ ratio of true labels $T$ and the estimated satisfaction $S$ as $F^{T(\alpha),S}$.
We exploit prevalent personalized recommendation methods as $F$, including LR~\cite{richardson2007predicting}, FM~\cite{rendle2010factorization}, and Wide\&Deep~\cite{cheng2016wide}.
All recommendation methods are implemented with the open-sourced code of Recbole~\cite{zhao2021recbole} and are applied with its default settings.  

\subsubsection{Protocols} 
We treat each video segment in the AMIGOS dataset as an item and assign it with one-hot encoding as item embedding.
For user embedding, the collected users' demographic information, personality profiles, and the mood~(PANAS) files are used and expressed as 71-dimensional embedding vectors.
We randomly split the user-item pairs into training, validating, and testing sets with a ratio of 8:1:1.
In the training set, we replace the true labels with the satisfactions inferred from brain signals, and we evaluate rating prediction performance with the true labels in the validating and testing set.

\section{results and discussions}
We empirically evaluate BTA in the satisfaction prediction task and utilize the estimated satisfaction to boost interactive information access performance to address the following research questions:
\begin{itemize}
	\item \textbf{RQ1} Can we effectively estimate satisfaction feedback signals~(e.g., feedback on search results and preference of items) with BTA from brain signals?
	\item \textbf{RQ2} Can we interactively provide satisfying information~(e.g., helpful search results and users’ preferred items) with the satisfaction inferred from the brain signals?
\end{itemize}
To address \textbf{RQ1}, we compare BTA with prevalent EEG classification baselines and delve into BTA's components in Section~\ref{Satisfaction Prediction}.
Then, we summarize the performance of the search result re-ranking task and the rating predicting tasks to answer \textbf{RQ2} in Section~\ref{Task 1: Search results re-ranking} and Section~\ref{Task 2: video rating prediction}, respectively.

\subsection{Satisfaction Prediction}
\label{Satisfaction Prediction}
\begin{table}[]

\caption{The performance of satisfaction estimation for each model. $*$ indicates difference compared to the BTA is significant with $p$-value $\textless 0.01$.} 
\begin{tabular}{lllll}
\toprule
\multirow{2}{*}{\textbf{Model}} & \multicolumn{2}{c}{\textbf{Search-Brainwave}} & \multicolumn{2}{c}{\textbf{AMIGOS}}     \\  
                       & F1             & AUC          & F1           & AUC      \\ \midrule
\multicolumn{3}{l}{\textbf{Topography-invariant}}          &              &           \\
DT                     & $0.5642^{*}$  & $0.5205^{*}$ & $0.5608^{*}$ & $0.6245^{*}$   \\
MLP                    & $0.6196^{*}$  & $0.5204^{*}$ & $0.5629^{*}$ & $0.6123^{*}$ \\
SVM                    & $0.6227^{*} $ & $0.5189^{*}$ & $0.5580^{*}$ & $0.5892^{*}$ \\ \midrule
\multicolumn{2}{l}{\textbf{CNN-based}}      &              &              &               \\
BENDR                  & $0.7118^{*} $ & $0.7291^{*}$ & $0.5580^{*}$ & $0.5869^{*}$ \\
EEGNet                 & $0.7254^{*}$  & $0.7614^{*}$ & $0.6025^{*}$& $0.6920^{*}$ \\
\midrule
\multicolumn{2}{l}{\textbf{GCN-based}}  &              &              &             \\
DGCNN                  & $0.7170^{*}$   & $0.7374^{*}$ & $0.6630^{*}$ & $0.7663^{*}$ \\
HetEmotionNet          & $0.7362^{*}$   & $0.7717^{*}$ & $0.6428^{*}$& $0.7405^{*}$ \\
RGNN                   & $0.7440^{*}$   & $0.7663^{*}$ & $0.6694^{*}$& $0.7782^{*}$  \\
\midrule
BTA~(ours)   & $\textbf{0.7837}$ & $\textbf{0.8278}$& $\textbf{0.7143}$& $\textbf{0.8353}$  \\\bottomrule
\end{tabular}
\label{Overall performance}
\vspace{-3mm}
\end{table}

\subsubsection{Overall performance}
Table~\ref{Overall performance} presents the satisfaction estimation performance in terms of F1-score and \ac{AUC}~\cite{powers2020evaluation} of different models.
From Table~\ref{Overall performance}, we have the following observations:
(1)~All \ac{CNN}-based models and \ac{GCN}-based models outperform topography-invariant models in most of the evaluation metrics.
The topography-invariant models directly concatenate all the \ac{EEG} channels' features together.
Thus the topographical information is omitted, which leads to limited performance.
(2)~In general, \ac{CNN}-based models perform worse than \ac{GCN}-based models.  
\ac{CNN}-based architectures compress the 3D topographical information into 2D representation and aggregate the information in adjacent \ac{EEG} channels.
Conversely, \ac{GCN}-based models utilize learnable adjacency matrixes to learn a more flexible aggregation strategy, which better exploits the topographical information than \ac{CNN}-based models.
(3)~The proposed BTA performs the best among all models. 
On the one hand, BTA introduces multi-centrality encoding to exploit the 3D spatial relations.
, while previous models don't take the 3D information into account
On the other hand, the spatial attention mechanism is applied to capture the topographical information adaptively.
Thus we can learn different aggregation strategies in different data samples.
Conversely, previous \ac{CNN}-based and \ac{GCN}-based models utilize common aggregation weights shared by all data samples.
One exception is HetEmotionNet, which uses mutual information between \ac{EEG} channels to learn the aggregation weight.
More detailed comparison on the exploitation of topographical information between BTA and HetEmotionNet is conducted in Section~\ref{Brain topography analysis}.

\subsubsection{Ablation study}
\begin{figure}[b]
    \hspace*{\fill}%
    \subcaptionbox{F1~(Search-Brainwave)}
    {\includegraphics[width=.491\linewidth]{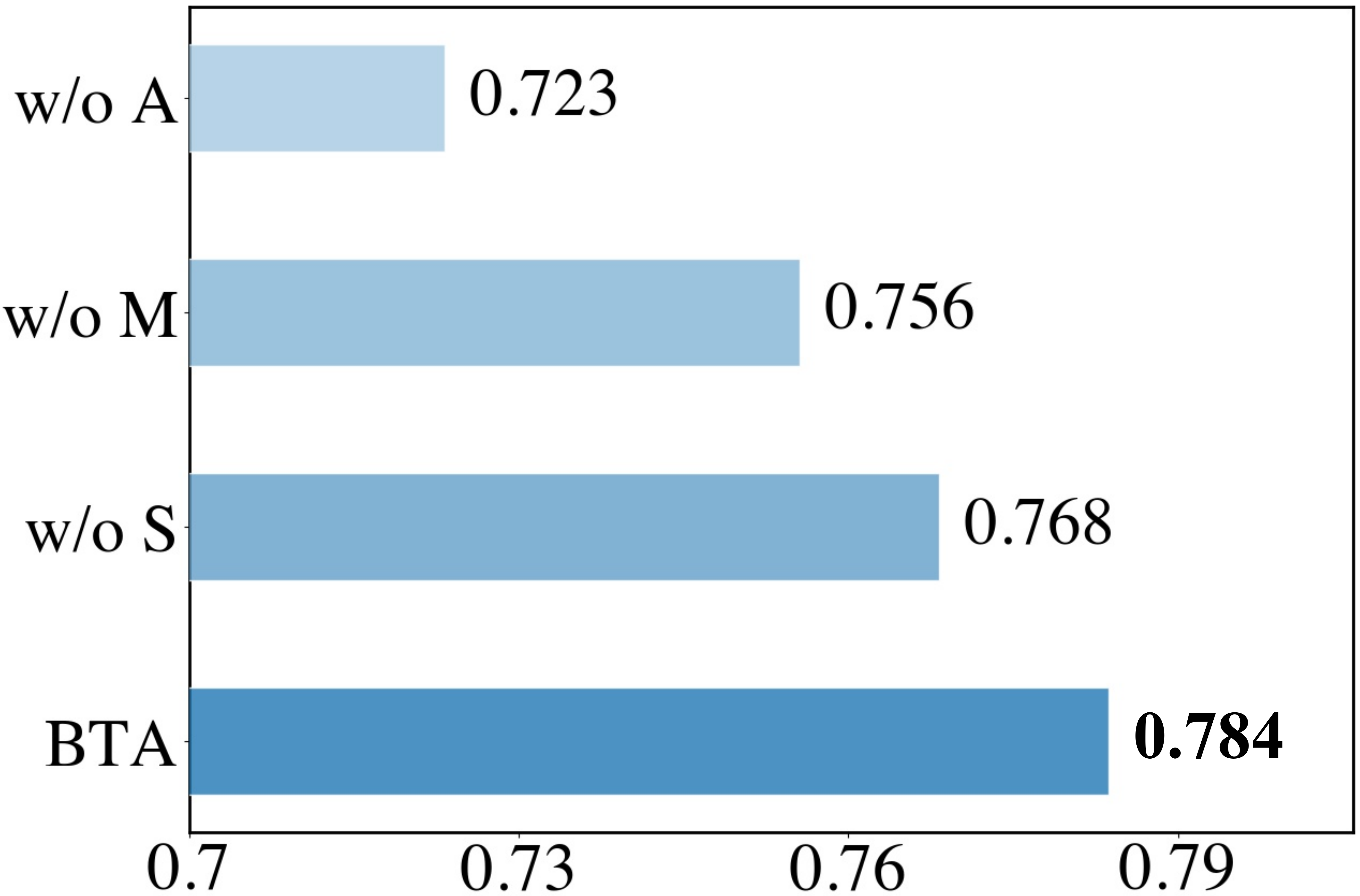}}
    \hfill\hfill\hfill\hfill%
    \subcaptionbox{F1~(AMIGOS)}
    {\includegraphics[width=.429\linewidth]{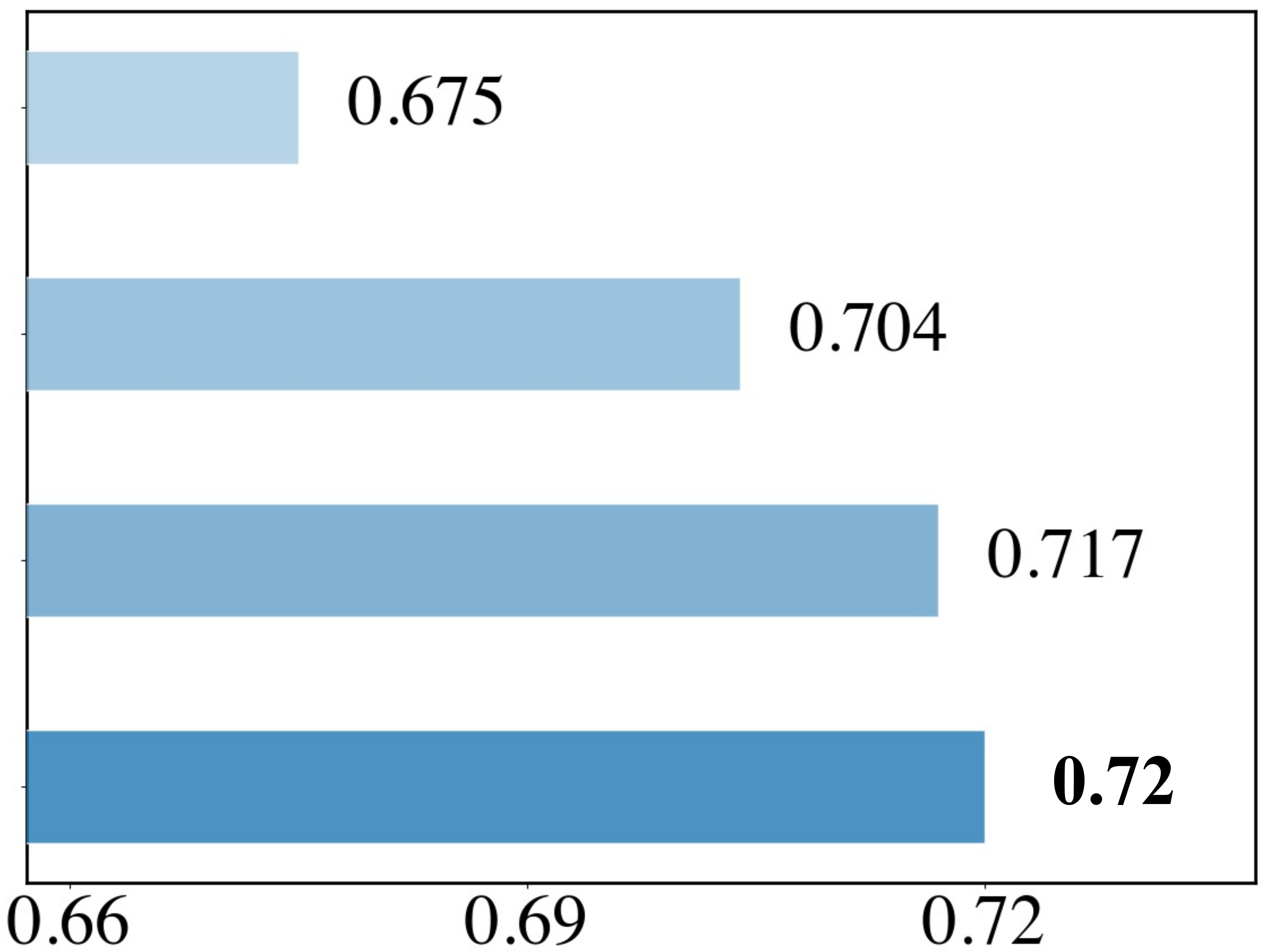}}%
    \hspace*{\fill}%
    \vspace{-2mm}
    \caption{Performance comparisons between BTA and its variants. A: spatial attention; M: multi-centrality encoding; S: unsupervised subtask.}
    \label{Ablation}
\end{figure}

To explore the effectiveness of different components in BTA, we compare it to a series of baselines without considering a variety of model structures or training strategies.
Figure~\ref{Ablation} shows the results of BTA and its variants.
w/o A, w/o M, and w/o S indicate the BTA model that masks the spatial attention module, the multi-centrality encoding module, and the subtask module, respectively.
As shown in Figure~\ref{Ablation}, there are different degrees of performance degradation in both tasks when masking one of the three components.
This implies that all of them facilitate the model performance.
Among these components, the spatial attention module plays the most important role, which suggests the effectiveness of adaptively aggregating the channel information.
Besides, the subtask module leads to the least performance improvement. 
Especially in the AMIGOS dataset, the performance improvement of BTA is merely 0.003 in terms of F1 compared to BTA w/o S.
We suggest that the AMIGOS dataset contains more training samples and supervised labels, and thus the unsupervised procedure to initialize the parameters is less necessary.

\newcommand{\tabincell}[2]{\begin{tabular}{@{}#1@{}}#2\end{tabular}}  
\begin{table*}[h]

\caption{A case study to investigate the rewriting performance. The bold words are related to the user intent.}
\label{case study}
\begin{tabular}{c|c|c|c}

\hlinew{0.8pt}
\textbf{Query} & \textbf{Search result/Satisfaction} & \textbf{ULM rewriting} & \textbf{SLM rewriting} \\ \hline
\tabincell{l}{permanent \\teeth}  & \tabincell{l}{(1) ..., online medical advice: dentist Mr.Song /\frownie{}  \\ (2) how \textbf{old} does a \textbf{child} grow its permanent teeth, ... /\smiley{}}
 & \tabincell{l}{permanent, teeth, dentist, \\ know, online, \textbf{child}, \textbf{kid}} & \tabincell{l}{permanent, teeth, \textbf{child},  \\ \textbf{old}, \textbf{when}, know, \textbf{kid}} \\ \hlinew{0.8pt}
\end{tabular}

\end{table*}

\subsubsection{Brain topography analysis}
\label{Brain topography analysis}
\begin{figure}[]
    \hspace*{\fill}%
    \subcaptionbox{BTA~(Satisfied)}
    {\includegraphics[height=2.5cm]{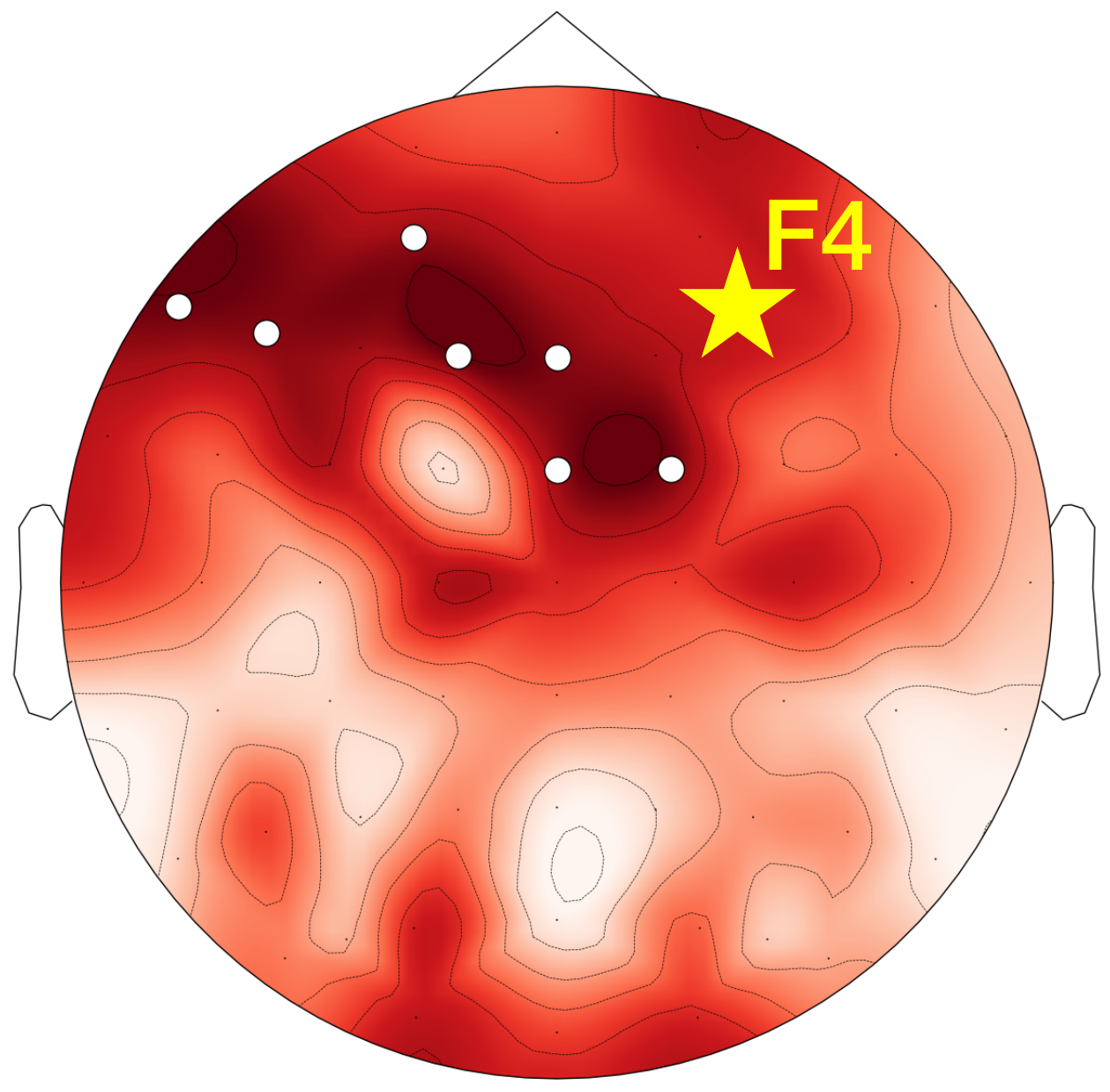}}
    \hfill\hfill\hfill\hfill%
    \subcaptionbox{BTA~(Unsatisfied)}
    {\includegraphics[height=2.5cm]{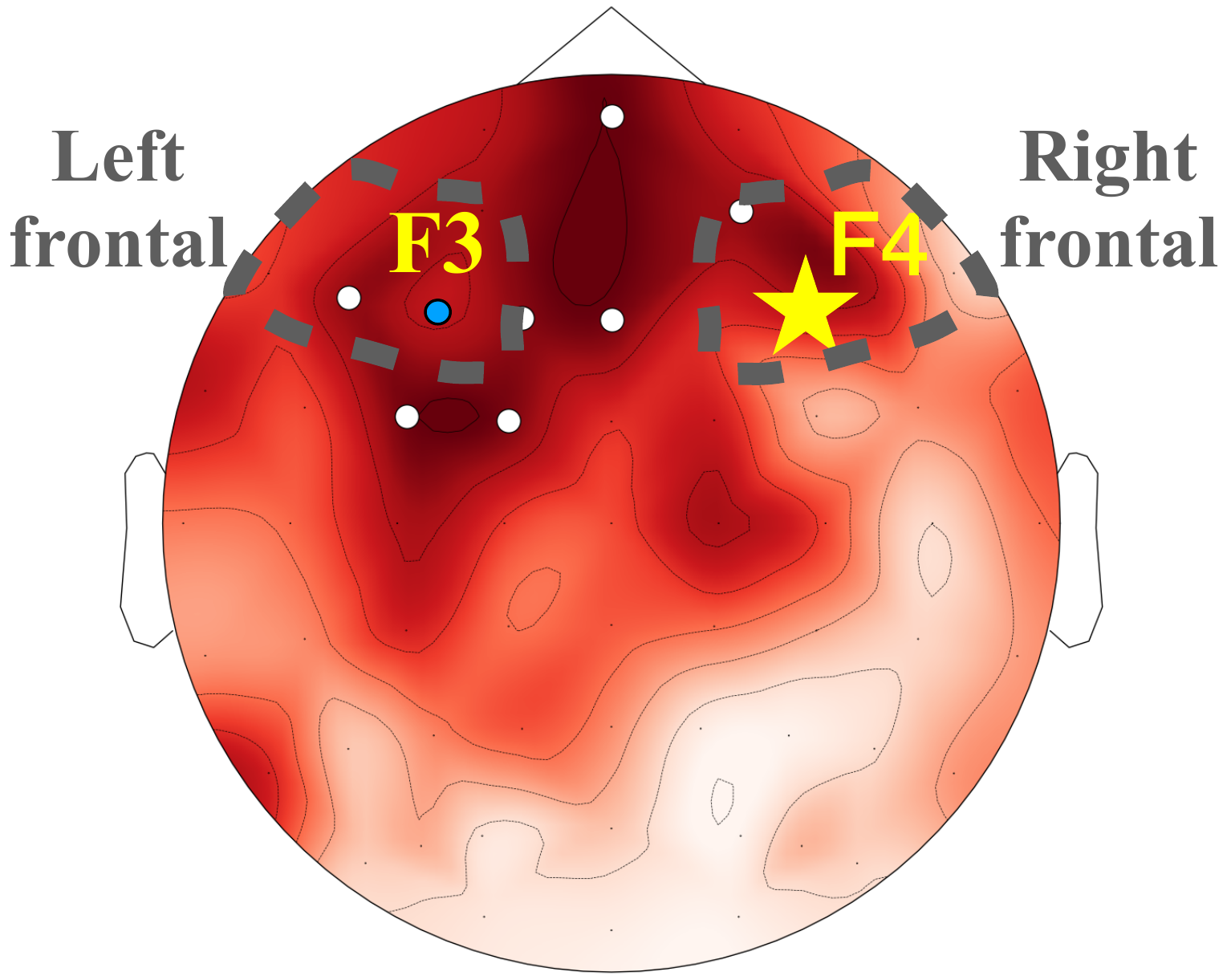}}
    \hfill\hfill\hfill\hfill%
    \subcaptionbox{Het~(Satisfied)}
    {\includegraphics[height=2.5cm]{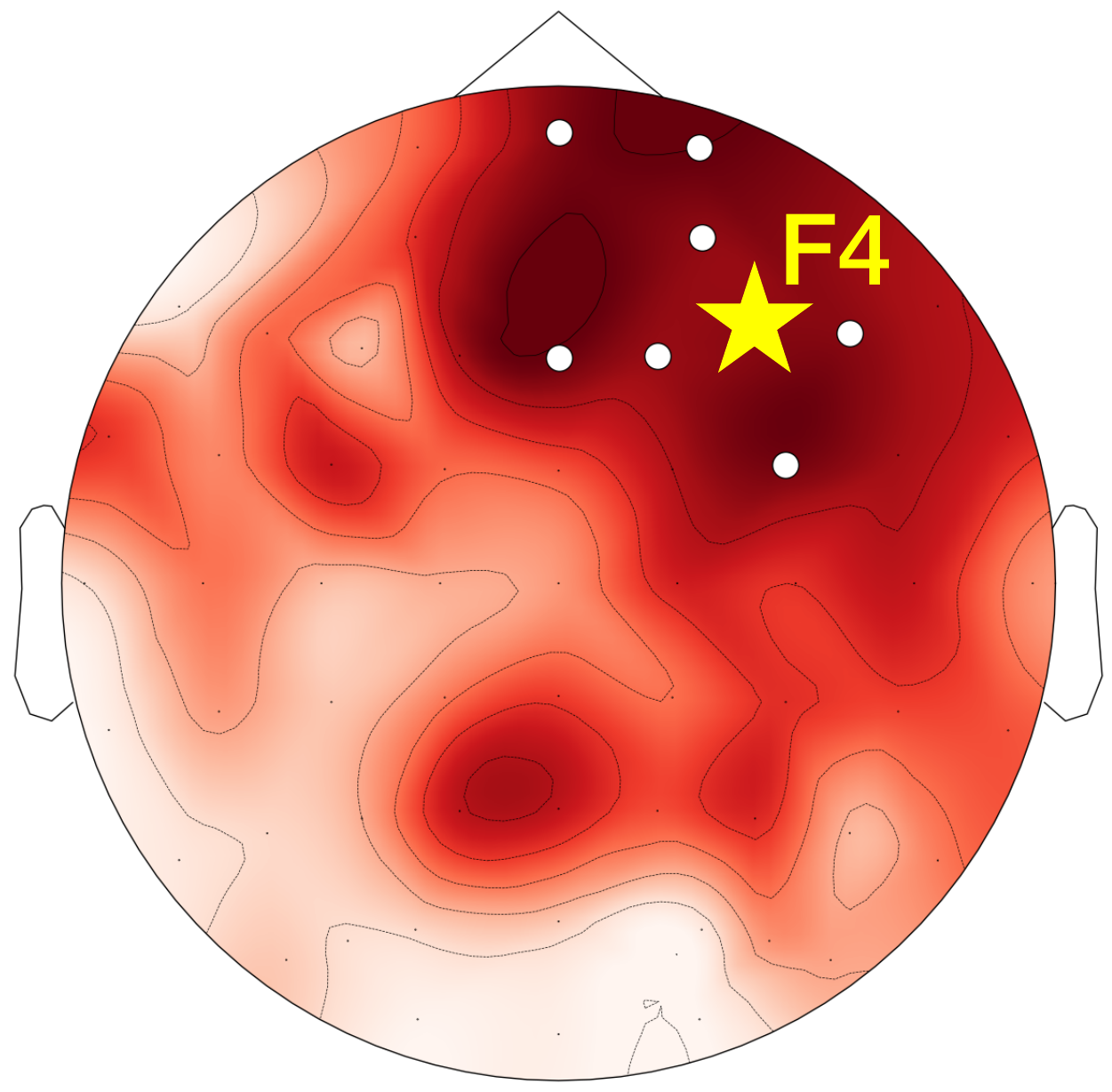}}%
    \hspace*{\fill}%
    \caption{The visualization of the aggregation weights in BTA/HetEmotionNet and satisfied/unsatisfied data samples. The darker color indicates higher aggregation weight to channel F4 and the highlighted channels have the highest weights among all channels. Het indicates HetEmotionNet.}
    \label{fig:aggregation}
    \vspace{-2mm}
\end{figure}
In this subsection, we explore the brain topographical information in our proposed model and the baseline HetEmotionNet by visualizing their aggregation weight.
Although BTA and HetEmotionNet have different architectures, both of their aggregation weights indicate how the model aggregate the information from other \ac{EEG} channels to a certain channel.
For BTA, the aggregation weight denotes the averaged attention weight between \ac{EEG} channels among all data samples.
For HetEmotionNet, the aggregation weight indicates the averaged edge weight~\cite{jia2021hetemotionnet} between \ac{EEG} channels among all data samples.

Figure~\ref{fig:aggregation} presents the visualization of the aggregation weight of various \ac{EEG} channels to channel F4 in the Search-Brainwave dataset.
Here we select channel F4 because previous neurological studies~\cite{schmidt2001frontal, aldayel2020deep} suggest that the frontal alpha asymmetry~(i.e., the $\alpha$ band difference between left and right frontal) indicates motivation, desire, and positive/negative feelings.
From Figure~\ref{fig:aggregation}, we can observe that the aggregation weight to channel F4 is higher in brain regions of right frontal and left frontal for HetEmotionNet and BTA, respectively.
HetEmotionNet adopts mutual information~\cite{kraskov2004estimating} to obtain the edge weight.
Since adjacent channels usually share higher mutual information scores, the aggregation process is just aggregating some of the most adjacent channels.
Conversely, BTA adaptively captures the topographical relationships between \ac{EEG} channels with attention mechanism and multi-centrality encoding.
Thus it aggregates information in a flexible manner.
Interestingly, the aggregation weights of BTA are higher in left frontal.
BTA utilizes a higher weight to capture the relationship between F4 and channels in left frontal, including F3, which implies BTA has good coherence to existing neurological studies~\cite{aldayel2020deep}. 
Besides, the satisfied and unsatisfied samples in BTA have different topographical relationships, which also demonstrate the data-dependent modeling strategy can be adaptive to different user statuses.

\subsection{Search result re-ranking performance}
\label{Task 1: Search results re-ranking}

\begin{table}[]
\caption{The performance of search result re-ranking for each model. $*$ indicates difference compared to SLM is significant with $p$-value $\textless 0.01$.}
\begin{tabular}{llllll}
\toprule
\textbf{Model} & NDCG@1  & NDCG@5 & NDCG@10 & MAP@10 \\
\midrule
BM25 & $0.6881^{*}$  & $0.7397^{*}$ & $0.8164^{*}$  & $0.7333^{*}$ \\
ULM & $0.7237^{*}$  & $0.7620^{*}$ & $0.8309^{*}$ & $0.7687^{*}$ \\
SLM   & $\textbf{0.7351}$ & $\textbf{0.7767}$ & $\textbf{0.8337}$   & $\textbf{0.7741}$ \\
\bottomrule
\end{tabular}

\label{tab:Search results re-ranking}
\end{table}

\begin{figure}[b]
\vspace{-3mm}

  \centering
  \includegraphics[width=0.75\linewidth]{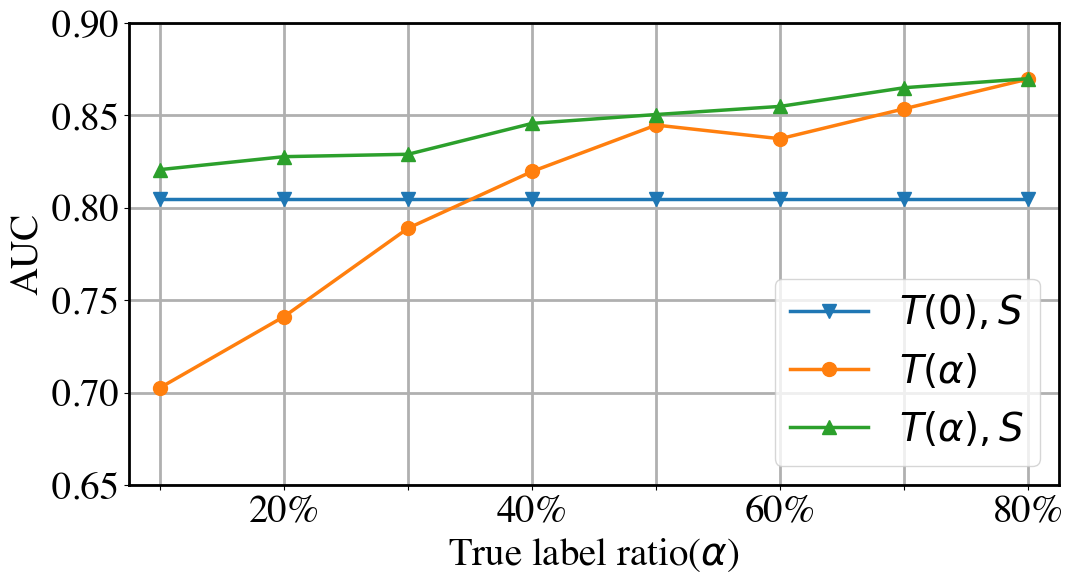}
  \caption{The performance of Wide\&Deep using only the estimated satisfaction $F^{T(0),S}$, different ratios of true labels $F^{T(\alpha)}$, and their combination $F^{T(\alpha),S}$.} 
  \label{fig:true}
  \Description[]{}
\end{figure}

Table~\ref{tab:Search results re-ranking} summarizes the performance of the search result re-ranking task.
We can observe that the proposed SLM outperforms BM25 and ULM.
It demonstrates the effectiveness of SLM, which rewrites the query with additional relevant words inferred from the search results with higher estimated user satisfaction. 
The significant improvements of SLM also demonstrate the benefits of introducing the estimated satisfaction into interactive search scenarios.

To investigate the rewriting performance of SLM and ULM, we conduct a case study to analyze their difference.
Table~\ref{case study} shows a search task that requires the user to explore information about when the period of growing permanent teeth is.
The ULM rewrites the query with words appeared in both search result 1~(e.g., ``online'', ``dentist'') and search result 2~(e.g., ``child'', ``kid'').
On the other hand, with the help of the estimated satisfaction, the SLM tends to rewrite the query with words that appeared in satisfying search results, i.e., search result two.
Therefore, words such as ``child'', ``old'', and ``when'' are used for rewriting in SLM, and the rewritten query can better present user intent.  
Then, the re-ranking performance can be improved with the rewritten query.

\begin{table}[t]
\caption{The AUC performance of personalized rating prediction for each model. $F^{T(0.1),S}$ indicates models using 0.1 ratio of true labels $T$ and the estimated satisfaction $S$.}
\begin{tabular}{llll}
\toprule
 \textbf{Model}     & LR    & FM &   Wide\&Deep  \\
\midrule
$F^{T(0.1)}$     & 0.6798    & 0.5010 & 0.7027\\
$F^{T(0.1),S}$   & \textbf{0.7537} & \textbf{0.8056} & \textbf{0.8207}\\
\bottomrule
\end{tabular}
\label{tab:video rating prediction}
\vspace{-3mm}
\end{table}

\subsection{Rating prediction performance}
\label{Task 2: video rating prediction}

Table~\ref{tab:video rating prediction} summarizes the rating prediction performances of different models using a true label ratio of 0.1~(see in Section~\ref{Methods_Task2}).
As shown in Table~\ref{tab:video rating prediction}, all recommendation models improve a large margin with the enhancement of the estimated satisfaction $S$.
This suggests the effectiveness of introducing brain signals into recommendation by utilizing the estimated satisfaction to train the model.

Additionally, to better understand to what extent brain signals can improve the interactive recommendation system, we explore the performance of Wide\&Deep using different ratios of true labels $T$ and whether using the estimated satisfaction $S$ or not.
Figure~\ref{fig:true} presents their performance comparison.
We observe that Wide\&Deep using only the satisfaction estimated with brain signals $S$~(i.e., $F^{T(0),S}$) are as effective as Wide\&Deep using about 30$\%$-40$\%$ true labels $T$~(i.e., $F^{T(\alpha)}$).
Besides, $F^{T(\alpha),S}$ outperforms $F^{T(\alpha)}$ at the same true label ratio $\alpha$, which indicates introducing the estimated satisfaction $S$ can boost the performance stably. 
Note that in recommendation scenarios, true user satisfaction is often scarce~\cite{wang2021denoising}.
Thus, the estimated satisfaction is valuable and here we demonstrate to what extent the satisfaction estimated with brain signals can improve the performance.

\section{Conclusion}
In this paper, we addresss a problem of utilizing brain signals for satisfaction modeling in interactive information systems.
Compared to conventional user signals, brain signals can directly present user status and thus contain less bias.
We then propose a Brain Topography Adaptive network~(BTA) to estimate user satisfaction with EEG signals.
BTA exploits the 3D topographical information of \ac{EEG} channels by multi-centrality encoding module and adaptively learns data-dependent aggregation strategies with spatial attention mechanisms.
Extensive experiments on the Search-Brainwave and the AMIGOS datasets demonstrate the outstanding performance of our model in comparison with various competitive baselines.
Moreover, to verify that the estimated satisfaction can help the interactive information access procedures, we conduct two downstream tasks in search and recommendation scenarios.
Experimental results show that brain signals can boost the system performance of result re-ranking in search and rating prediction in recommendation. 

With wearable devices becoming more portable and cheaper, information systems are possible to collect users' psychophysiological signals in situations such as \ac{VR} applications~\cite{baumgartner2006neural} and the disabled service~\cite{chen2022web}.
As BCIs become more prevalent and cheaper, we believe that more application scenarios will soon emerge.
Besides directly controlling information systems, we reveal another benefit of BCI for information systems.
We suggest utilizing brain signals to automatically predict user satisfaction with well-designed models, and then, the system can better understand the user and interactively provide useful information.
Future studies may include interactive systems in broad situations, an overall framework for \ac{EEG}-enhanced search systems, and online learning algorithms for real-life \ac{EEG} based satisfaction estimation.

\section{Acknowledgement}
This work is supported by the Natural Science Foundation of China~(Grant No. 61732008), Beijing Academy of Artificial Intelligence~(BAAI), and Tsinghua University Guoqiang Research Institute.

\bibliographystyle{ACM-Reference-Format}

\balance
\bibliography{references}
\newpage
\section{SUPPLEMENTARY MATERIAL}

\subsection{Dataset}
\label{Dataset}
\subsubsection{Search-Brainwave dataset}

The Search-Brainwave dataset~\cite{sigir2022ye} records the brain signals generated by 18 participants when performing search tasks. 
Each participant averagely accomplishes 69.6 search tasks and examines 3.6 corresponding search results in a search task.
During the procedures, their explicit feedback~(i.e., satisfied or dissatisfied) to search results are collected.
And the dataset also provides the true relevance label of each search result for performance evaluation.
We choose this dataset because it is a public benchmark specially designed for interactive search tasks.

For feature extraction, we utilize the officially preprocessed temporal and spectral features~(i.e., raw signals and \ac{DE}~\cite{duan2013differential} features, respectively) and treat the brain signals in response to each search result as a data sample.

\subsubsection{AMIGOS dataset}
The AMIGOS dataset~\cite{miranda2018amigos} is a public dataset that includes EEG, electrocardiogram~(ECG), and other psychological signals generated by 40 participants under video stimulations.
Each participant watches 16 short videos and 4 long videos in two experiments. 
They rate each video in ``valence'', ``arousal'', ``dominance'', ``familiarity'', and ``liking'' from 1 to 9.
Although the AMIGOS dataset is not specially designed for recommendation scenarios, we choose this dataset for recommendation scenario because it contains rich participant profiles~(anonymized participants' data, personality profiles, and mood~(PANAS) profiles).
The user profile can be utilized as user embeddings for the presonalized rating prediction task~(see in Section~\ref{problem:Rating Prediction}).
Conducting a specially designed user study on interactive recommendation scenarios is left as future work.

In our experiments, we adopt the data in short videos experiment since the long videos experiment is conducted in different configurations.
We divide the ``liking'' annotation into satisfied and unsatisfied with a threshold of five for our satisfaction estimating tasks.
For feature extraction, we utilize the officially preprocessed \ac{EEG} signals and apply a non-overlapping window with a length of one second to divide each video into several segments.
The window length is different from the original paper~\cite{miranda2018amigos} but agrees with a lot of existing \ac{EEG} classification tasks~\cite{song2018eeg, jia2021hetemotionnet}.
We utilize a different time window since the original paper also considers other physiological signals, such as ECG, which requires a longer time window than EEG~\cite{pyakillya2017deep}.
For brain signals in response to each video segment, the raw signals are used as temporal features, and we use Fourier transform over four frequency bands~(i.e., $\theta, \alpha, \beta, \gamma$) to extract the \ac{DE}~\cite{duan2013differential} features as spectral features. 

\subsection{Parameter Setups of Satisfaction Prediction Baselines}
For the satisfaction prediction baselines of various EEG classification models~(see in Section~\ref{baselines}), we use the original parameter settings according to their original papers and released codes~\cite{song2018eeg,jia2021hetemotionnet,zhong2020eeg,lawhern2018eegnet,kostas2021bendr}.
However, there exist some implementation details as follows.
For EEGNet~\cite{lawhern2018eegnet}, the original parameter settings are not suitable for AMIGOS dataset due to the difference in input data length.
Thus we decrease the second convolution layer's kernel size to 7 and the last pooling layer's kernel size to $(2,1)$ for experiments on AMIGOS dataset.
For RGNN~\cite{zhong2020eeg}, the released code excludes the regularized parameters, and we empirically set it as 0.001.
The implementation code of our experiment is publicly available in https://github.com/YeZiyi1998/DL4EEG-Classification.

\vspace{-1.5mm}
\begin{algorithm}[b]
\caption{The Training Procedures of BTA.}
\label{algorithm1}
\LinesNumbered
\KwIn{User's brain signals in response to search results $\langle X^1, ... ,X^{N} \rangle$; Usefulness of search results $\langle Y^1, ... ,Y^{N} \rangle$; The Initialize BTA model $\Phi$.} 
\BlankLine
Generate time domain and spectral domain features $X^t=\langle x^{1,t}, ..., x^{N,t} \rangle$ and $X^s=\langle x^{1,s}, ..., x^{N,s} \rangle$.

$\Phi^{'} = Copy(\Phi)$.

\For{iteration=1,2,...}{
\For{ all $X$ in $\langle X^1, ... ,X^{N} \rangle$}{
Generate randomized binary noise masks $W_{t,mask}$ and $W_{s,mask}$.

$\widetilde{X} = W_{mask} \odot X, X \in \{X_t, X_s\}$.

Compute reconstruction loss $L_{MSE}=L_{MSE}(\Phi^{'},\widetilde{X},X)$.

Update $\Phi^{'}$ with $L_{MSE}$.
}
}

Replace the centrality embedding vectors of $\Phi$ with $\Phi^{'}$.

\For{iteration=1,2,...}{
\For{ all $X$ in $\langle X^1, ... ,X^{N} \rangle$ and $Y$ in $\langle Y^1, ... ,Y^{N} \rangle$}{

Compute classification loss $L=L(\Phi,X,Y)$.

Update $\Phi$ with $L$.
}
}

\vspace{0.2mm}

return $\Phi$\;
\end{algorithm}
\vspace{-1.5mm}

\end{document}